\documentclass[aps,prc,superscriptaddress,twocolumn]{revtex4}

\usepackage{amssymb}
\usepackage{graphicx}
\usepackage{epsfig}%
\usepackage{bm}
\usepackage{hyperref}
\usepackage{mathrsfs}
\usepackage{multirow}
\usepackage{footmisc}
\usepackage[normalem]{ulem}
\usepackage{url}

\begin{document}

\title{Effect of energy conservation on in-medium $\mbox{NN}\to \mbox{N}\Delta$ cross section in isospin-asymmetric nuclear matter}


\author{Ying Cui}
\email{yingcuid@163.com}
\affiliation{China Institute of Atomic Energy, Beijing 102413, China}

\author{Yingxun Zhang}
\email{zhyx@ciae.ac.cn}
\affiliation{China Institute of Atomic Energy, Beijing 102413, China}
\affiliation{Guangxi Key Laboratory Breeding Base of Nuclear Physics and Technology, Guangxi Normal University, Guilin 541004, China}

\author{Zhuxia Li}
\affiliation{China Institute of Atomic Energy, Beijing 102413, China}

\begin{abstract}
In this paper, the in-medium $NN\rightarrow N\Delta$ cross section is calculated in the framework of the one-boson exchange model by including the isovector mesons, i.e. $\delta$ and $\rho$ mesons.
Due to the isospin exchange in the $NN\rightarrow N\Delta$ process, the vector self-energies of the outgoing particles are modified relative to the incoming particles in isospin asymmetric nuclear matter, and it leads to the effective energies of the incoming $NN$ pair being different from the outgoing $N\Delta$ pair. This effect is investigated in the calculation of the in-medium $NN\rightarrow N\Delta$ cross section. With the corrected energy conservation, the cross sections of the $\Delta^{++}$ and $\Delta^+$ channels are suppressed, and the cross sections of the $\Delta^0$ and $\Delta^-$ channels are enhanced relative to the results obtained without properly considering the potential energy changes. Our results further confirm the dependence of medium correction factor, $R=\sigma_{ NN\rightarrow N\Delta}^*/\sigma_{NN\rightarrow N\Delta}^{\text{free}}$, on the charge state of $NN\rightarrow N\Delta$ especially around the threshold energy, but the isospin splitting of medium correction factor $R$ becomes weak at high beam energies.
\end{abstract}

\date{\today}


\pacs{Valid PACS appear here}

\maketitle

\section{Introduction}
The isospin dependence of an in-medium nucleon-nucleon ($NN$) cross section is the subject of much interest in the field of intermediate energy neutron-rich heavy ion collisions (HIC). As a key ingredient of the transport model\cite{Bertsch88,Aichelin91}, the magnitudes of the in-medium $NN$ cross sections largely influence the frequency of nucleon-nucleon collisions, which provides the short-range repulsion and competes with the nucleonic mean field potential in heavy ion collisions. And thus, the different medium correction on the $NN$ cross sections can influence the predictions of HIC observables, such as stopping power\cite{JYLiu01, zhang07,lehaut10,lopez14}, collective flow\cite{zhang07,lwchen00,YJWang16}, isospin transport\cite{BALi05,Zhang12},  nuclear reaction cross section\cite{ouli08}, and the understanding of the reaction mechanism.

Currently, one of the hot debates in the field of HICs is the constraint of symmetry energy at suprasaturation density by comparing the $\pi^-/\pi^+$ ratio data from the FOPI Collaboration \cite{Resid07} with calculations from transport models.
Very different conclusions are inferred from  different models. For example, the isospin-dependent Boltzmann-Uehling-Uhlenbeck (IBUU04)\cite{Xiao2009prl} and the isospin-dependent Boltzmann-Langevin (IBL)\cite{Xie2013} calculations favor the super soft symmetry energy, but the calculation from the improved isospin dependent quantum molecular dynamics model (ImIQMD)\cite{Feng2010} needs a super stiff symmetry energy to reproduce the data. The calculations from the relativistic Vlasov-Uehling-Uhlenbeck (RVUU)\cite{Song2015} and the T\"{u}bingen quantum molecular dynamics model (TuQMD)\cite{Cozma2017} support the symmetry energy between super soft and super stiff by considering the medium threshold effects, corrected energy conservation, and pion potential. However, the calculations with the Boltzmann-Uehling-Uhlenbeck model from Michigan State University (pBUU)\cite{Hong2014} show the pion yield ratio is not sensitive to the symmetry energy by including the strong pion interaction.

This divergence has stimulated a lot of works to understand pion production and propagation in HICs, such as the threshold effect of $\Delta$ production \cite{Ferini2005}, which is caused by the different potential energies between the incoming and the outgoing colliding pairs in isospin asymmetric nuclear matter and the pion optical potential \cite{Hong2014,ZhenZhang2017}, which is caused by the pion self-energy. For reproducing the pion multiplicity data, the in-medium $NN\rightarrow N\Delta$ cross section ($\sigma^*_{NN\rightarrow N\Delta}$) \cite{Ferini2005,Song2015} is needed. The issues of Pauli blocking, cluster formations \cite{Natsu16}, energy conservation issues \cite{Cozma17}, and the strength of the $\Delta$ symmetry potential\cite{BaoanliPRC2015} in the dynamics of the simulation are also investigated.
Among all those factors, the in-medium $NN\rightarrow N\Delta$ cross section is one of the key ingredients for the $\pi-N-\Delta$ loop in the simulations because it will directly influence the first $\Delta$ production which can decay into nucleon and pion or rescatter with nucleons. Most of the transport codes adopted the free space $NN\rightarrow N\Delta$ cross section, i.e., the $\sigma_{NN\rightarrow N\Delta}^{\text{free}}$ taken from Ref. \cite{Huber1994}, or phenomenological in-medium cross section, i.e.,  $\sigma_{NN\rightarrow N\Delta}^{*}=R \sigma_{NN\rightarrow N\Delta}^{\text{free}}$,  in the collision integral of transport models\cite{Song2015} where the medium correction factor $R$ is independent of the channels of the $NN\rightarrow N\Delta$ process. For reproducing the pion yield, $R<1$ is required, and it is consistent with what one found in the theoretical calculations on $\sigma^*_{NN\rightarrow N\Delta}$ \cite{Larionov2003} from the one-pion exchange model.

However, for the further studies in this field, it is required to include the isospin dependent medium correction factor $R$ as well as the considering of the isovector mean field potential in transport models. It stimulates theoretical studies of the isospin dependent in-medium correction factor. Recently, the isospin dependence of elementary two-body  cross section, i.e., $\tilde{\sigma}_{NN\to N\Delta}^* $, was studied in the framework of relativistic Boltzmann-Uehling-Uhlenbeck (RBUU) microscopic transport theory by Li and Li in Ref.~\cite{QingfengLi2017}.
Their results showed  $\tilde{\sigma}^*_{NN\rightarrow N\Delta}$ has a sharp increment around the threshold energy without considering the $\Delta$ mass distribution, and the medium correction factor $R$ obviously depends on the isospin channels of $NN\rightarrow N\Delta$, i.e., $pp\rightarrow n\Delta^{++}$, $pp\rightarrow p\Delta^{+}$, $pn\rightarrow n\Delta^{+}$, $pn\rightarrow p\Delta^{0}$, $nn\rightarrow n\Delta^{0}$, and $nn\rightarrow p\Delta^{-}$, in isospin asymmetric nuclear matter. As a short living resonance, the $\Delta$ subsequently decays into a nucleon and a pion.
Thus, the measured cross section for $NN\rightarrow N\Delta$ is the elementary two-body cross section averaged over the mass distribution of $\Delta$ resonance, and the medium correction factor $R$ in the transport models also contains the effect of the $\Delta$ mass distribution.
Furthermore, the isospin exchange process modifies the scalar and vector self-energies of incoming and outgoing channels in the $NN\rightarrow N\Delta$ process in isospin asymmetric nuclear matter, and it results in the difference in the effective energies between incoming and outgoing particles. It requires that one has to properly consider the energy conservation in the calculation of $\sigma^*_{NN\rightarrow N\Delta}$ as well as for the $\Delta$ threshold energy\cite{Ferini2005,Song2015,ZhenZhang2017}, but few theoretical works on the in-medium $NN\rightarrow N\Delta$ cross section considered it.

In this paper, we study the in-medium $NN\rightarrow N\Delta$ cross sections in isospin asymmetric nuclear matter by considering the corrected energy conservation in the one-boson exchange model\cite{Huber1994,Machleidt1987} which is described in Sec.~\ref{model}.
In the model we used, the isovector-scalar $\delta$ and the isovector-vector $\rho$ mesons are included for describing the isospin asymmetric nuclear matter and the in-medium $NN\rightarrow N\Delta$ cross section. In Sec.~\ref{ec}, the energy conservation issue caused by the $\rho$ meson is discussed in the calculation of the in-medium $NN\to N\Delta$ cross section. In Sec.~\ref{xs}, we present the results of the in-medium $NN\rightarrow N\Delta$ cross section in isospin asymmetric nuclear matter, and a summary is given in Sec.~\ref{summary}.

\section{The Model}
\label{model}
\subsection{In-medium $NN\rightarrow N\Delta$ cross section}
For the calculation of the $NN\rightarrow N\Delta$ cross section in isospin asymmetric nuclear matter, we use the one-boson exchange model with the relativistic Lagrangian which includes the nucleon and  $\Delta$ degree with the Rarita-Schwinger spinor of the spin-3/2 \cite{Huber1994,Machleidt1987,Benmerrouche1989} coupling to $\sigma$, $\omega$, $\rho$, $\delta$, and $\pi$ mesons.
This theoretical framework is similar to the work in\cite{Larionov2003}, and the main difference between two works is the form of effective Lagrangian. In this work, the $\delta$ and $\rho$ mesons are included for describing isospin asymmetric nuclear matter and the isospin dependence of the $NN\rightarrow N\Delta$ cross sections which were not considered in the symmetric condition from Ref. \cite{Larionov2003}.
\begin{figure}[htbp]
\begin{center}
    \includegraphics[scale=0.12]{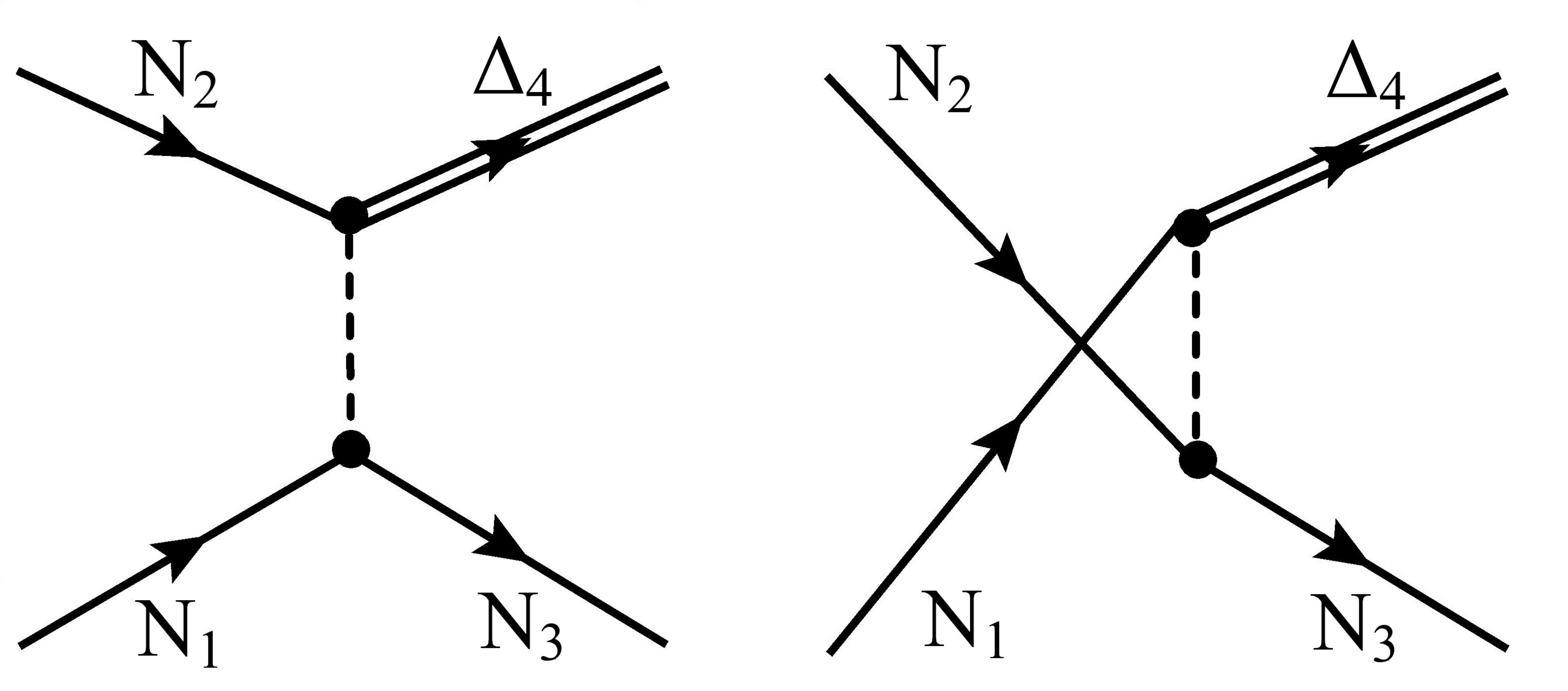}
    \caption{The left diagram is the direct term and the right is the exchange term of the Feynmann diagram.}\label{fig0}
\end{center}
\end{figure}
The Lagrangian we used is as follows:
\begin{equation}
\label{Lag}	
\mathcal{L}=\mathcal{L}_I+\mathcal{L}_F,
\end{equation}
where $\mathcal{L}_F$ is,
\begin{eqnarray}
\label{lag_f}
\mathcal{L}_{F}=&&\bar{\Psi}[i\gamma_{\mu}\partial^{\mu}-m_{N}]\Psi+\bar{\Delta}_{\lambda}[i\gamma_{\mu}\partial^{\mu}-m_{\Delta}]\Delta^{\lambda}\\
&&+\frac{1}{2}\partial_{\mu}\sigma\partial^{\mu}\sigma-\frac{1}{2}m^{2}_{\sigma}\sigma^{2}-\frac{1}{3}g_{2}\sigma^{3}-\frac{1}{4}g_{3}\sigma^{4}\nonumber\\
&&-\frac{1}{4}\omega_{\mu\nu}\omega^{\mu\nu}+\frac{1}{2}m^{2}_{\omega}\omega_{\mu}\omega^{\mu}+\frac{1}{2}\left(\partial_{\mu}\bm{\pi}\partial^{\mu}\bm{\pi}-m^{2}_{\pi}\bm{\pi}^{2}\right)\nonumber\\
&&-\frac{1}{4}\bm{\rho}_{\mu\nu}\bm{\rho}^{\mu\nu}+\frac{1}{2}m^{2}_{\rho}\bm{\rho}_{\mu}\bm{\rho}^{\mu}+\frac{1}{2}\left(\partial_{\mu}\bm{\delta}\partial^{\mu}\bm{\delta}-m^{2}_{\delta}\bm{\delta}^{2}\right), \nonumber
\end{eqnarray}
and $\mathcal{L}_I$ is
\begin{eqnarray}
\label{lag_i}
\mathcal{L}_I&=&\mathcal{L}_{NN}+\mathcal{L}_{\Delta \Delta}+\mathcal{L}_{N\Delta}\nonumber\\
&=&g_{\sigma NN}\bar{\Psi}\Psi\sigma-g_{\omega NN}\bar{\Psi}\gamma_{\mu}\Psi\omega^{\mu}-g_{\rho NN}\bar{\Psi}\gamma_{\mu}\bm{\tau} \cdot\Psi\bm{\rho}^{\mu}\nonumber\\
&&+\frac{g_{\pi NN}}{m_{\pi}}\bar{\Psi}\gamma_{\mu}\gamma_{5}\bm{\tau} \cdot\Psi\partial^{\mu}\bm{\pi}+g_{\delta NN}\bar{\Psi}\bm{\tau} \cdot\Psi\bm{\delta}\nonumber\\
&&+g_{\sigma \Delta \Delta}\bar{\Delta}_{\mu}\Delta^{\mu}\sigma-g_{\omega \Delta \Delta}\bar{\Delta}_{\mu}\gamma_{\nu}\Delta^{\mu}\omega^{\nu} \nonumber\\
&&-g_{\rho \Delta\Delta}\bar{\Delta}_{\mu}\gamma_{\nu}\bm{\mbox{T}} \cdot\Delta^{\mu}\bm{\rho}^{\nu}+\frac{g_{\pi \Delta\Delta}}{m_{\pi}}\bar{\Delta}_{\mu}\gamma_{\nu}\gamma_{5}\bm{\mbox{T}} \cdot\Delta^{\mu}\partial^{\nu}\bm{\pi}\nonumber\\
&&+g_{\delta \Delta\Delta}\bar{\Delta}_{\mu}\bm{\mbox{T}} \cdot\Delta^{\mu}\bm{\delta}+\frac{g_{\pi N\Delta}}{m_{\pi}}\bar{\Delta}_{\mu}\bm{\mathcal{T}}\cdot \Psi\partial^{\mu}\bm{\pi}\nonumber\\
&&+\frac{ig_{\rho N\Delta}}{m_{\rho}}\bar{\Delta}_{\mu}\gamma_{\nu}\gamma_{5}\bm{\mathcal{T}}\cdot \Psi\left(\partial^{\nu}\bm{\rho}^{\mu}-\partial^{\mu}\bm{\rho}^{\nu}\right)+h.c. ~
\end{eqnarray}
$\omega_{\mu\nu}$ and $\bm{\rho}_{\mu\nu}$ in Eq. (\ref{lag_f}) are defined by $\partial_{\mu}\omega_{\nu}-\partial_{\nu}\omega_{\mu}$ and $\partial_{\mu}\bm{\rho}_{\nu}-\partial_{\nu}\bm{\rho}_{\mu}$, respectively. Here $\bm{\tau}$ and $\mathbf{T}$ are the isospin matrices of the nucleon and $\Delta$ \cite{Machleidt1987,Benmerrouche1989}, and $\bm{\mathcal{T}}$ is the isospin transition matrix between the isospin 1/2 and the 3/2 fields\cite{Huber1994}.

In the quasiparticle approximation \cite{Baym76}, the in-medium cross sections are introduced via the replacement of the vacuum plane waves of the initial and final baryons by the plane waves obtained by the solution of the nucleon and $\Delta$ equation with the scalar and vector fields. In detail, the matrix elements $\mathcal{M}^*$ for the inelastic scattering process $NN\rightarrow N\Delta$ are obtained by replacing the nucleon and $\Delta$ masses and momenta in free space with their effective masses and kinetic momenta \cite{Larionov2003}, i.e., $m^*=m+\Sigma^S$ and $p^{*\mu}=p^{\mu}-\Sigma^\mu$, where $\Sigma^S$ and $\Sigma^{\mu}$ are the scalar and vector self-energies.

The Feynmann diagrams corresponding to the inelastic-scattering $NN\rightarrow N\Delta$ process are shown in Fig.\ref{fig0}, which include the direct and exchange processes. The $\mathcal{M}^*$ matrix for the interaction Lagrangian Eq.(\ref{lag_i}) can be written by the standard procedure \cite{Huber1994},
\begin{equation}
\mathcal{M}^*=\mathcal{M}_d^{*\pi}-\mathcal{M}_e^{*\pi}+\mathcal{M}_d^{*\rho}-\mathcal{M}_e^{*\rho}
\end{equation}
where
\begin{eqnarray}
\mathcal{M}_d^{*\pi}&=&-i\frac{g_{\pi NN} g_{\pi N\Delta} I_d  }{ m_{\pi}^{2}( Q^{*2}_{d}- m_{\pi}^{2})}[\bar{\Psi}(p_3^* ) \gamma_{\mu}\gamma_5  Q_{d}^{*\mu}  \Psi(p_1^*)]\nonumber\\
&&\times[\bar{\Delta}_{\nu} (p_4^* ) Q_d^{*\nu}  \Psi(p_2^* )]\\
\mathcal{M}_d^{*\rho}&=&i\frac{g_{\rho NN} g_{\rho N\Delta}I_d}{m_{\rho} }[\bar{\Psi}(p_3^* ) \gamma_{\mu} \Psi(p_1^* )]\\\nonumber
&&\times\frac{g^{\mu\tau}-Q_d^{*\mu} Q_d^{*\tau}/m^{2}_{\rho}}{Q_d^{*2}-m^{2}_{\rho}}\\\nonumber
&&\times[\bar{\Delta}_{\sigma} (p_4^* ) \gamma_{\lambda} \gamma_{5} (Q_d^{*\lambda} \delta_{\sigma\tau}-Q_d^{*\sigma} \delta_{\lambda\tau}) \Psi(p_2^* )]
\end{eqnarray}

The upper index in $\mathcal{M}^{*\text{meson}}_{d,e}$ refers to the exchanged boson and the lower index represents the direct or exchange process, and $Q_{d}^{*\mu}=p_{3}^{*\mu}-p_{1}^{*\mu}$ is for the direct term. The exchange term $\mathcal{M}^*_e$ is obtained by $p_{1}^{*\mu}\longleftrightarrow p_{2}^{*\mu}$ and $Q_{e}^{*\mu}=p_{3}^{*\mu}-p_{2}^{*\mu}$. The isospin factors $I_d$, $I_e$, and the spin projection matrix for the spin-3/2 particles can be found from  Ref.\cite{Huber1994}.

The in-medium $NN\rightarrow N\Delta$ cross section can be written as
\begin{eqnarray}
&&\sigma^*_{NN\rightarrow N\Delta}=\int_{m^*_{\Delta,\text{min}}}^{m^*_{\Delta,\text{max}}} dm^*_{\Delta}f(m^*_{\Delta})\tilde{\sigma}^*(m^*_{\Delta}).
\label{eq:xsnd1}
\end{eqnarray}
$\tilde{\sigma}^*( m^*_{\Delta})$ is the in-medium elementary two-body cross section, and $f(m^*_{\Delta})$
is the  mass distribution of $\Delta$ resonance. In the center-of-mass frame of colliding nucleons, the in-medium elementary two-body cross section reads
\begin{eqnarray}
\label{eq:xsnd2}
\tilde{\sigma}^*( m^*_{\Delta})
&=&\frac{1}{4F^*}\int
\frac{d^3 \textbf{p}_3^*}{(2\pi)^3 2E_3^* }  \frac{d^3 \textbf{p}_4^*}{(2\pi)^3 2E_4^* }\\\nonumber
&&(2\pi)^4\delta^{4}(p_1+p_2-p_3-p_4)\overline{|\mathcal{M}^*|^2}\\\nonumber
&=&\frac{1}{64\pi^2}\int \frac{|\textbf{p}^{*}_{\text{out, c.m.}}|}{\sqrt{s^*_{\text{in}}}\sqrt{s^{* }_{out}}|\textbf{p}^{*}_{\text{in, c.m.}}|} \overline{|\mathcal{M}^*|^2}  d\Omega,
\end{eqnarray}
where $\overline{|\mathcal{M}^*|^2}=\frac{1}{(2s_{1}+1)(2s_2+1)}\sum\limits_{s_{1}s_{2}s_{3}s_{4}}|\mathcal{M}^*|^2$,
$\textbf{p}^{*}_{\text{in, c.m.}}$ and $\textbf{p}^{*}_{\text{out, c.m.}}$ are the center-of-mass momenta of the incoming (1 and 2) and outgoing particles (3 and 4), respectively. $F^*=\sqrt{(p^*_{1}p^*_{2})^2-p^{*2}_{1}p^{*2}_{2}}=\sqrt{s^*_{\text{in}}}|\textbf{p}^{*}_{\text{in, c.m.}}|$ is the invariant flux factors, $s^*_{\text{in}}=(p^*_1+p^*_2)^2$, and $s^*_{\text{out}}=(p^*_3+p^*_4)^2$. Here, one should note that the crucial requirement of the two-body collisions is the energy-momentum conservation in terms of the incoming and outgoing canonical momenta ($p^\mu_{1,2}$, $p^\mu_{3,4}$), i.e., $\delta^4(p_1+p_2-p_3-p_4)$ instead of the kinetic momentum $p^{* \mu}$. This will be discussed in more detail below in Sec.~\ref{ec}.

\subsection{The coupling constants used in the model}
In our paper, we take the NL$\rho\delta$ parameter set given in Ref. \cite{Liu2005} by which the properties of isospin asymmetric nuclear equation of state \cite{Dutra14} can be well reproduced.
The values of nuclear matter parameters, such as $K_0=240$ MeV, $m^*/m=0.75$, $S_0=30.6$  MeV, and $L=101.46$ MeV and the values of $S_0$ and $L$ are close to the symmetry energy constraints from the neutron-proton differential flow data \cite{Russto}. The coupling constant $g_{\pi N\Delta}$ is determined by analyzing the $\Delta$-isobar decay width from \cite{Dmitriev1986}.
Concerning the coupling constant $g_{\rho N\Delta}$, in this paper we use the value derived from the static quark model \cite{Engel90, Huber1994} $g_{\rho N\Delta}\approx\frac{\sqrt{3}}{2} g_{\rho NN} \frac{m_{\rho}}{m_N}$. All the values of these parameters are listed in the second and third column of Table~\ref{table-para}.
For the coupling constant parameters of $g_{m\Delta\Delta}$, $m=\sigma, \omega, \rho, \delta$, we simply take $g_{m\Delta\Delta}=g_{mNN}$ as the same as that did in many studies with the transport models \cite{Song2015,QingfengLi2017,Larionov2003}.

The form factors are adopted for effectively considering the contribution from the high-order terms and  finite size of the baryons \cite{Huber1994,Vetter1991}, which read,
\begin{equation}
\label{factorn}
F_N (t^*)=\frac{\Lambda_N^2}{\Lambda_N^2-t^*}  exp\left(-b\sqrt{s^*-4m_N^{* 2}}\right)
\end{equation}
\begin{equation}
\label{factord}
F_{\Delta}(t^*)=\frac{\Lambda_{\Delta}^2}{\Lambda_{\Delta}^2-t^*}.
\end{equation}
Here, $F_N (t^* )$ is the form factor for the nucleon-meson-nucleon coupling, $b$=0.046 GeV$^{-1}$ for both $\rho NN$ and $\pi NN$, and $\Lambda_{\rho NN}\approx\Lambda_{\pi NN}=\Lambda_{N}$=1.0 GeV.
$F_\Delta (t^*)$ is the form factor for $\Delta$ and the cutoff parameter $\Lambda_{\Delta}$$\approx0.41$ GeV for both $\rho N\Delta$ and $\pi N\Delta$ which is determined by  best fitting the data of the $NN\rightarrow N\Delta$ cross section in free space \cite{Baldini1987} ranging from $\sqrt{s}$=2.0 to 5.0 GeV.

\begin{table}[htbp]
\begin{center}
\caption{Parameters in the effective Lagrangian, $m_{\sigma}$=0.550, $m_{\omega}$=0.783, $m_{\rho}$=0.770, $m_{\delta}$=0.980, $m_{\pi}$=0.138, $m_{N}$=0.939, $m_{0,\Delta}$=1.232 GeV , $g_{2}/g_{\sigma NN}^3$=0.03302 fm$^{-1}$, $g_{3}/g_{\sigma NN}^4$=-0.00483. The coupling constants $g_{mNN}$ and $g_{mN\Delta}$ are dimensionless.}\label{table-para}

\begin{tabular}{c|c|c|c}
  \hline
  \hline
   Meson & $g_{mNN}$ & $g_{mN\Delta}$ & $g_{m\Delta\Delta}/g_{mNN}$ \\
    \hline
   $\sigma$ & 8.9679 &  & 1  \\
    $\omega$& 9.2408 &  & 1 \\
   $\rho$& 6.9256 & 4.9183 & 1 \\
    $\delta$ & 7.8525 &   & 1 \\
   $\pi$ &1.008 &  2.202 & 1 \\
   \hline
   \hline	 		 	 	 			 	 	
\end{tabular}
\end{center}
\end{table}

\section{Energy CONSERVATION IN  $\mathbf{\sigma^*_{NN\rightarrow N\Delta}}$}
\label{ec}
For the $NN\to N\Delta$ process, the energy-momentum conservation is in terms of the incoming and outgoing canonical momenta ($p^\mu_{1,2}$, $p^\mu_{3,4}$), i.e., $\delta^4(p_1+p_2-p_3-p_4)$, whatever they are in the symmetric or asymmetric nuclear medium. In the symmetric nuclear medium, it can also be fulfilled by simply using the effective mass ($m^*$) and kinetic momentum ($p^{*\mu}$), because the self-energies of the incoming and the outgoing particles are the same. However, in the isospin asymmetric nuclear medium, the scalar and vector self-energies between the incoming and the outgoing channels may be different and thus the energy conservation should exactly use the canonical momenta instead of the kinetic momenta $p^{*\mu}$. This idea was first proposed by Ferini \textit{et al.} in Ref. \cite{Ferini2005} for calculating its effects on the threshold energy for $NN\to N\Delta$ and followed by other works  \cite{Song2015,ZhenZhang2017} in the transport models. However, for calculating the in-medium $NN\to N\Delta$ cross section, the same effects should also be considered simultaneously with that of isospin splitting effects. The details of this effect are given in the following.

Since all the calculations performed in this paper are in the center of mass of the colliding particles, it coincides with the nuclear matter rest frame where the effective momentum $\textbf{p}_i^*=\textbf{p}_i$ due to the vanishment of the spatial components of the vector field, i.e., $\mathbf{\Sigma}=0$. The effective energy reads as
\begin{equation}
p_i^{*0}=p^{0}_{i}-\Sigma^{0}_{i}
\end{equation}
and
\begin{equation}
\Sigma^{0}_{i}=g_{\omega NN}\bar{\omega}^{0}+g_{\rho NN}t_{3,i}\bar{\rho}^{0}_3.
\end{equation}
Here $t_{3,i}$ represents the third component of the isospin of the nucleon and $\Delta$, where $t_{3,n}=-1$, $t_{3,p}=1$, $t_{3,\Delta^{++}}=1$, $t_{3,\Delta^{+}}=\frac{1}{3}$, $t_{3,\Delta^{0}}=-\frac{1}{3}$, $t_{3,\Delta^{-}}=-1$, and $\bar{\rho}^{0}_3=\frac{g_{\rho NN}}{m^2_\rho}(\rho_{p}-\rho_{n}$).

For symmetric nuclear matter, the energy conservation $p^{0}_1+p^{0}_2=p^{0}_3+p^{0}_4$ is equal to $p^{*0}_1+p^{*0}_2=p^{*0}_3+p^{*0}_4$. This is because the scalar and vector self-energies between the incoming and the outgoing particles are the same, i.e. $\Sigma_1^0+\Sigma_2^0=\Sigma_3^0+\Sigma_4^0$ and $\Sigma_1^S+\Sigma_2^S=\Sigma_3^S+\Sigma_4^S$ (or $\Delta\Sigma^{0}=\Sigma^{0}_1+\Sigma^{0}_2-\Sigma^{0}_3-\Sigma^{0}_4=0$, $\Delta\Sigma^{S}=\Sigma^{S}_1+\Sigma^{S}_2-\Sigma^{S}_3-\Sigma^{S}_4=0$) due to $\bar{\rho}^{0}_3=0$.  Thus, using the kinetic momentum, i.e., $\delta^4 (p^*_1+p^*_2-p^*_3 -p^*_4)$, in the formula of cross section Eq.~(\ref{eq:xsnd2}) can fulfill the energy momentum conservation.

In isospin asymmetric nuclear matter, the scalar and vector self-energies of the incoming and outgoing particles may differ as shown in Table~\ref{table1}.
For example, in the case of  $pp\rightarrow  n \Delta^{++}$, $\Sigma^{0}_p+\Sigma^{0}_p\ne \Sigma^{0}_n+\Sigma^{0}_{\Delta^{++}}$, i.e., $\Delta \Sigma^0\neq0$, as shown in the first row of Table \ref{table1}.  $\Delta\Sigma^S$ and $\Delta\Sigma^0$ have the opposite contributions to the energy of particles, in which $\Delta\Sigma^S$=29.3 and $\Delta\Sigma^0=-39.8$ MeV at the normal density.
Consequently, $p^{*0}_1+p^{*0}_2$ differs from $p^{*0}_3+p^{*0}_4$, and $s^*_{\text{in}}\ne s^*_{\text{out}}$ in Eq.~(\ref{eq:xsnd2}). $s^*_{\text{in}}$ and $s^*_{\text{out}}$ are related according to the following relationship,
\begin{equation}
\label{sinstar}
\sqrt{s^*_{\text{in}}}+\Sigma^0_{N_1}+\Sigma^0_{N_2}=\sqrt{s^*_{\text{out}}}+\Sigma^0_{N_3}+\Sigma^0_{\Delta_4}.
\end{equation}
It is derived from
\begin{eqnarray}
\label{eq:sin}
s&=&(p_{N_{1}}+p_{N_{2}})^2\nonumber\\
&=&(\sqrt{m^{*2}_{N_{1}}+\mathbf{p}^{*2}_{N_{1}}}+\sqrt{m^{*2}_{N_{2}}+\mathbf{p}^{*2}_{N_{2}}}+\Sigma^0_{N_1}+\Sigma^0_{N_2})^2\nonumber\\
& &-(\mathbf{p}^*_{N_{1}}+\mathbf{p}^*_{N_2})^2\\\nonumber
&=&(p_{N_{3}}+p_{\Delta_4})^2
\end{eqnarray}
in the center-of-mass frame, where $\mathbf{p}^*_{N_1}=-\mathbf{p}^*_{N_2}$ and $\mathbf{p}^*_{N_3}=-\mathbf{p}^*_{\Delta_4}$.
Thus, using $\delta^4 (p^*_1+p^*_2-p^*_3 -p^*_4)$ can not be equivalent to the energy-momentum conservation in  isospin asymmetric nuclear matter. Properly imposing the energy-momentum conservation is to use the canonical momentum, i.e., $\delta^4 (p_1+p_2-p_3 -p_4)$, in the formula of in Eq.(\ref{eq:xsnd2}) can fulfill the energy-momentum conservation.

In our paper, we use the energy conservation factor in the in-medium cross section calculations as $\delta (p^{0}_1+p^{0}_2-p^{0}_3 -p^{0}_4)=\delta (p^{*0}_1+p^{*0}_2-p^{*0}_3 -p^{*0}_4+\Delta\Sigma^0)$ which is  the same idea as in the study of threshold effects \cite{Ferini2005,Song2015,ZhenZhang2017}.
We named this corrected energy conservation ``EC-C" in the following text. In order to understand its effect, we also show the results of the $NN\rightarrow N\Delta$ cross section obtained by considering the kinetic momentum conservation, i.e., $\delta (p^{*0}_1+p^{*0}_2-p^{*0}_3 -p^{*0}_4)$, and we named it as ``EC-K" in the following.
\begin{table}[htbp]
\begin{center}
\small{\caption{Difference between the initial and the final scalar and
vector mean fields in the $NN\to N\Delta$ process as well as in the decay of $\Delta$ resonances ($\Delta \rightarrow N+\pi$).
$\Delta\Sigma^S=\Sigma^{S}_{1}+\Sigma^{S}_{2}-\Sigma^{S}_{3}-\Sigma^{S}_{\Delta} $, $\Delta\Sigma^0=\Sigma^{0}_{1}+\Sigma^{0}_{2}-\Sigma^{0}_{3}-\Sigma^{0}_{\Delta} $, $\Delta\Sigma_d^S=\Sigma^S_\Delta-\Sigma^S_N$, and $\Delta\Sigma_d^0=\Sigma^0_\Delta-\Sigma^0_N$. All entries are in MeV. It is similar to Table II of Ref. \cite{Song2015}.}\label{table1}
\begin{tabular}{c|cc|cc}
  \hline
 Scattering & $\Delta\Sigma^S$ & $\Delta\Sigma^S(\rho_{0})$ & $\Delta\Sigma^0$ & $\Delta\Sigma^0(\rho_{0})$  \\
  \hline
  $pp\to n\Delta^{++}$ &$-2g_{\delta NN}\bar{\delta}_3$ & 29.3 & $2g_{\rho NN}\bar{\rho}^{0}_3$ &$-39.8$\\

 $pp\to p\Delta^{+}$ &$-\frac{2}{3}g_{\delta NN}\bar{\delta}_3$ &9.8& $\frac{2}{3}g_{\rho NN}\bar{\rho}^{0}_3$ & $-13.3$ \\

 $np \to n\Delta^{+}$ &$-\frac{2}{3}g_{\delta NN}\bar{\delta}_3$ &9.8& $\frac{2}{3}g_{\rho NN}\bar{\rho}^{0}_3$ & $-13.3$\\

  $np \to p\Delta^{0}$ &$\frac{2}{3}g_{\delta NN}\bar{\delta}_3$ &$-9.8$& $-\frac{2}{3}g_{\rho NN}\bar{\rho}^{0}_3$ &13.3\\

  $nn\to n \Delta^{0}$ &$\frac{2}{3}g_{\delta NN}\bar{\delta}_3$ &$-9.8$& $-\frac{2}{3}g_{\rho NN}\bar{\rho}^{0}_3$ &13.3 \\

  $nn\to p \Delta^{-}$ &$2g_{\delta NN}\bar{\delta}_3$ &$-29.3$& $-2g_{\rho NN}\bar{\rho}^{0}_3$ &39.8 \\
  \hline
  \hline
  Decay & $\Delta\Sigma^{S}_{d}$ & & $\Delta\Sigma^{0}_{d}$ & \\
  \hline

  $\Delta^{++}\to p\pi^{+}$ & 0 & 0 & 0 &0\\

 $\Delta^{+}\to p\pi^{0}$ &$\frac{2}{3}g_{\delta NN}\bar{\delta}_3$ &$-9.8$& $-\frac{2}{3}g_{\rho NN}\bar{\rho}^{0}_3$ &13.3 \\

 $\Delta^{+}\to n\pi^{+}$ &$-\frac{4}{3}g_{\delta NN}\bar{\delta}_3$ &19.6 & $\frac{4}{3}g_{\rho NN}\bar{\rho}^{0}_3$ &$-26.6$\\

 $\Delta^{0}\to p\pi^{-}$ &$\frac{4}{3}g_{\delta NN}\bar{\delta}_3$ &$-19.6$ &$-\frac{4}{3}g_{\rho NN}\bar{\rho}^{0}_3$ &26.6 \\

  $\Delta^{0}\to n\pi^{0}$ &$-\frac{2}{3}g_{\delta NN}\bar{\delta}_3$ &9.8& $\frac{2}{3}g_{\rho NN}\bar{\rho}^{0}_3$ &$-13.3$\\

  $\Delta^{-}\to n\pi^{-}$ &0 & 0& 0 & 0 \\
  \hline
  \hline
  \end{tabular}}
\end{center}
\end{table}

The minimum $\Delta$ mass $m^*_{\Delta,\text{min}}$ in the formula of the cross section is determined by  $\Delta \rightarrow N+ \pi$ in isospin asymmetric nuclear matter as in Ref. \cite{ZhenZhang2017} when both $N$ and $\pi$ are at rest, and the modification of the scalar and vector self energies in this isospin exchange process should also be considered. Thus,
$m^*_{\Delta,\text{min}}=m^*_{N}+\Sigma^{0}_{N}+m^*_\pi+\Pi_P(\omega, \mathbf{q})-\Sigma^{0}_{\Delta}$=$m^*_{N}+m^*_\pi-\Delta\Sigma_d^0$, with $\Delta\Sigma^0_d=\Sigma^{0}_{N}+\Pi_P(\omega, \mathbf{q})-\Sigma^{0}_{\Delta}$. Considering the $m^*_{\pi}/m_{\pi}$ is less than $\sim$10\% at the normal density from the calculations by Kaiser and Weise \cite{Kaiser01} in isospin asymmetric nuclear matter, we neglect the medium effect on the pion's mass and simply take $m^*_{\pi}=m_{\pi}$ in this paper. Thus, we have $\Delta\Sigma_d^0=\Sigma_\Delta^0-\Sigma_{N}^0$. The values of $\Delta\Sigma_d^0$ and $\Delta\Sigma^S_d=\Sigma^S_\Delta-\Sigma^S_N$ at the normal density are also listed in Table~\ref{table1}.
The maximal $\Delta$ mass $m^*_{\Delta,\text{max}}$ is evaluated based on the Eq.(\ref{eq:sin}) for producing  $N$ and $\Delta$ at rest, and it leads to
\begin{equation}
m^*_{\Delta,\text{max}}=\sqrt{s}-m^*_{N_{3}}-\Sigma^0_{N_{3}}-\Sigma^0_{\Delta_4}.
\end{equation}

The in-medium $\Delta$ mass distribution $f(m^*_\Delta)$ is another important ingredient of the in-medium $NN\rightarrow N\Delta$ cross section for which the proper energy conservation is also necessary since  $f(m^*_\Delta)$ is related to the $\Delta\rightarrow N+\pi$ process. In this paper, the spectral function of  $\Delta$ is taken as in Ref. \cite{Larionov2003},
\begin{equation}
\label{eq:bt}
f(m^*_{\Delta})=\frac{2}{\pi}\frac{m^{* 2}_{\Delta}\Gamma(m^{*}_{\Delta})}{(m^{*2}_{0,\Delta}-m^{*2}_{\Delta})^2+m^{*2}_{\Delta}\Gamma^2(m^{*}_{\Delta}) }.
\end{equation}
Here, $m^*_{0,\Delta}$ is the effective pole mass of $\Delta$. The decay width $\Gamma(m^*_\Delta)$ is taken as a parameterization form \cite{Larionov2003}
\begin{eqnarray}
\label{eq:gama}
\Gamma(m^{*}_{\Delta})&=&\Gamma_{0}\frac{q^{3}(m^{* }_{\Delta},m^*_N,m^*_\pi)}{q^{3}(m^{*}_{0,\Delta},m^*_N,m^*_\pi)}\\\nonumber
&&\times\frac{q^{3}(m^{*}_{0,\Delta},m^*_N,m^*_\pi)+\eta^2}{q^{3}(m^{* }_{\Delta},m^*_N,m^*_\pi)+\eta^2}\frac{m^{*}_{0,\Delta}}{m^{*}_{\Delta}}
\end{eqnarray}
where
\begin{eqnarray}
\label{eq:qm123}
&&q(m^{*}_\Delta,m^*_{N},m^*_\pi)=\\\nonumber
&&\sqrt{\frac{[(m^*_\Delta+\Sigma^{0}_{\Delta}-\Sigma^{0}_{N})^2+m_{N}^{*2}-m_{\pi}^{* 2}]^2}
{4(m^{*}_\Delta+\Sigma^{0}_{\Delta}-\Sigma^{0}_{N})^2}-m_{N}^{*2}}
\end{eqnarray}
is the center-of-mass momentum of the nucleon and pion from the decay of $\Delta$ in its rest frame. The factor of $(m^*_\Delta+\Sigma^{0}_{\Delta}-\Sigma^{0}_{N})$ in Eq.~(\ref{eq:qm123}) comes from properly considering the energy conservation in the $\Delta\rightarrow N\pi$ process in the isospin asymmetric nuclear matter
\footnotetext[1] {In the rest frame of $\Delta$ in the isospin asymmetric nuclear matter,
$\sqrt{s}=(m^*_\Delta+\Sigma^0_\Delta)=\sqrt{m^{*2}_N+\textbf{q}^2}+\Sigma^0_N+\sqrt{m^{*2}_\pi+\textbf{q}^2}+\Pi_P(\omega,\textbf{q})$.
In our approach, we assume the pion mass and momentum are not affected by the nuclear mean field.
}\footnotemark[1].
The coefficients of $\Gamma_0$=0.118 GeV and $\eta$=0.2 GeV/$c$ are used in the above the parameterization formula.

\section{Results and discussion}
\label{xs}

\subsection{$\mathbf{N}$ and $\mathbf{\Delta}$ effective masses in isospin asymmetric matter}
\begin{figure}[htbp]
\begin{center}
    \includegraphics[scale=0.28]{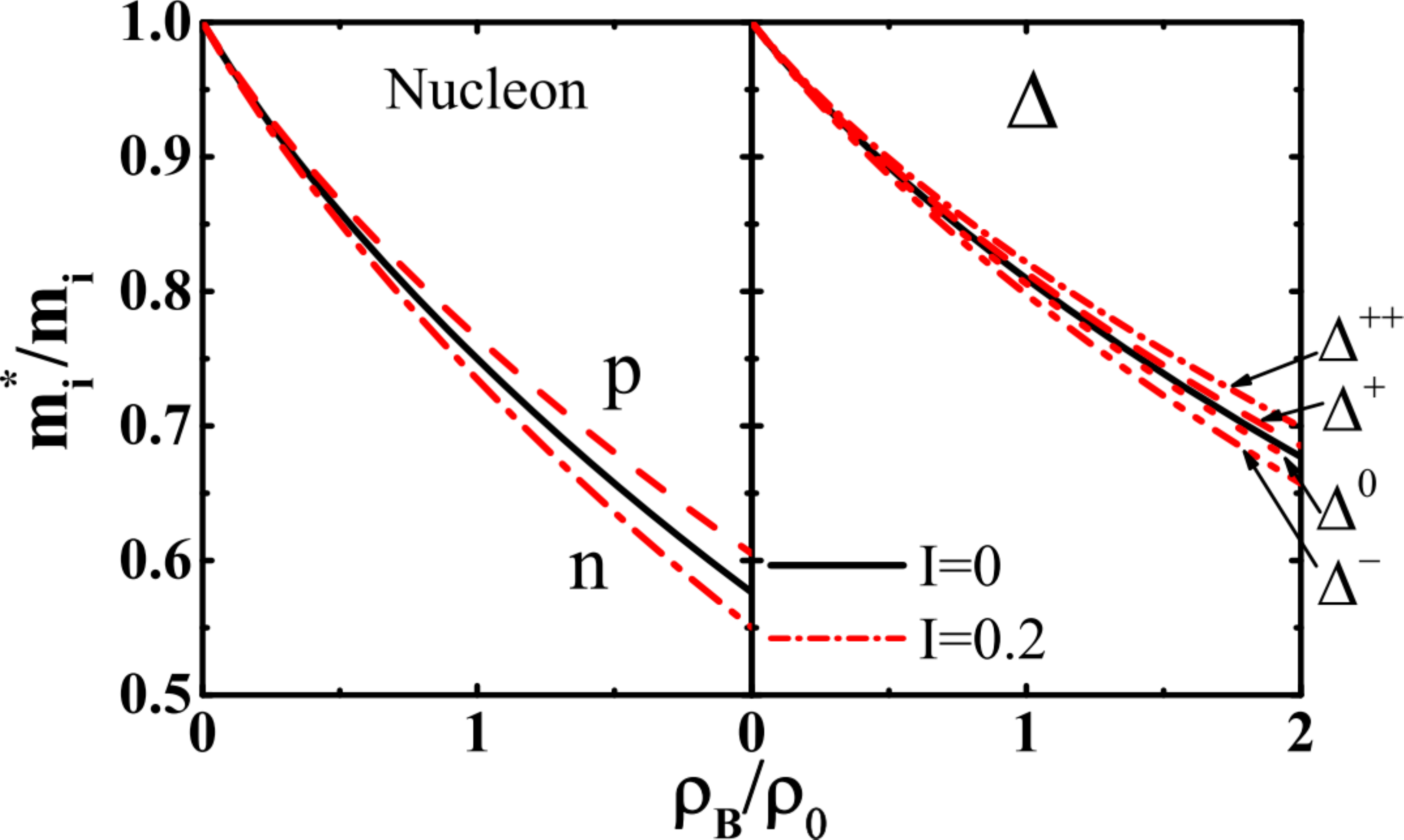}
    \caption{(Color online) The effective masses of the nucleon  and the effective pole masses of $\Delta$ as a function of $\rho_B/\rho_0$. The black solid lines are for symmetric matter $I=0$, and the red dashed or red dotted lines are for $I=0.2$. }\label{fig1}
\end{center}
\end{figure}

The Dirac effective masses of nucleons and $\Delta$'s are calculated in the relativistic mean-field approximation; they read
\begin{equation}
m^{*}_{i}=m_{i}+\Sigma^{S}_{i},
\label{eq:efmnd}
\end{equation}
where
\begin{equation}
\Sigma^{S}_{i}=-g_{\sigma NN}\bar{\sigma}- g_{\delta NN}t_{3,i}\bar{\delta}_3,
\label{eq:efmnd2}
\end{equation}
and $\bar{\delta}_3=\frac{g_{\delta NN}}{m^2_{\delta}}(\rho^{S}_p-\rho^{S}_n)$.
Figure~\ref{fig1} presents the Dirac effective masses for the nucleon and $\Delta$ in the symmetric nuclear matter (black solid lines) and in the neutron-rich matter with an isospin asymmetry $I=(\rho_n-\rho_p)/\rho_B=0.2$ (red dashed and dotted lines), where $\rho_B=\rho_n+\rho_p$.
The left panel shows the effective masses for the nucleons, and the right panel is for the effective $\Delta$ pole masses.
In symmetric nuclear matter, there is no effective mass splitting for nucleons and $\Delta$'s, and $m_i^*/m_i$=0.75 with $i=n, p$, $m_i^*/m_i$=0.81 with $i=\Delta^{++}$, $\Delta^+$, $\Delta^0$ and $\Delta^-$ at the saturation density with the parameter set in Table~\ref{table-para}.
In the neutron-rich matter, the effective masses of the nucleons and $\Delta$'s are split due to the contributions from the isovector-scalar $\delta$ meson. There is  $m_p^*>m_n^*$, $m_{0,\Delta^{++}}^*>m_{0,\Delta^{+}}^*>m_{0,\Delta^{0}}^*>m_{0,\Delta^{-}}^*$ in the neutron-rich matter. The splitting magnitudes of the effective masses for nucleons and $\Delta$'s depend on the coupling constant $g_{\delta NN}$.

\subsection{Cross section and its medium correction}
\begin{figure}[htbp]
\begin{center}
    \includegraphics[scale=0.33]{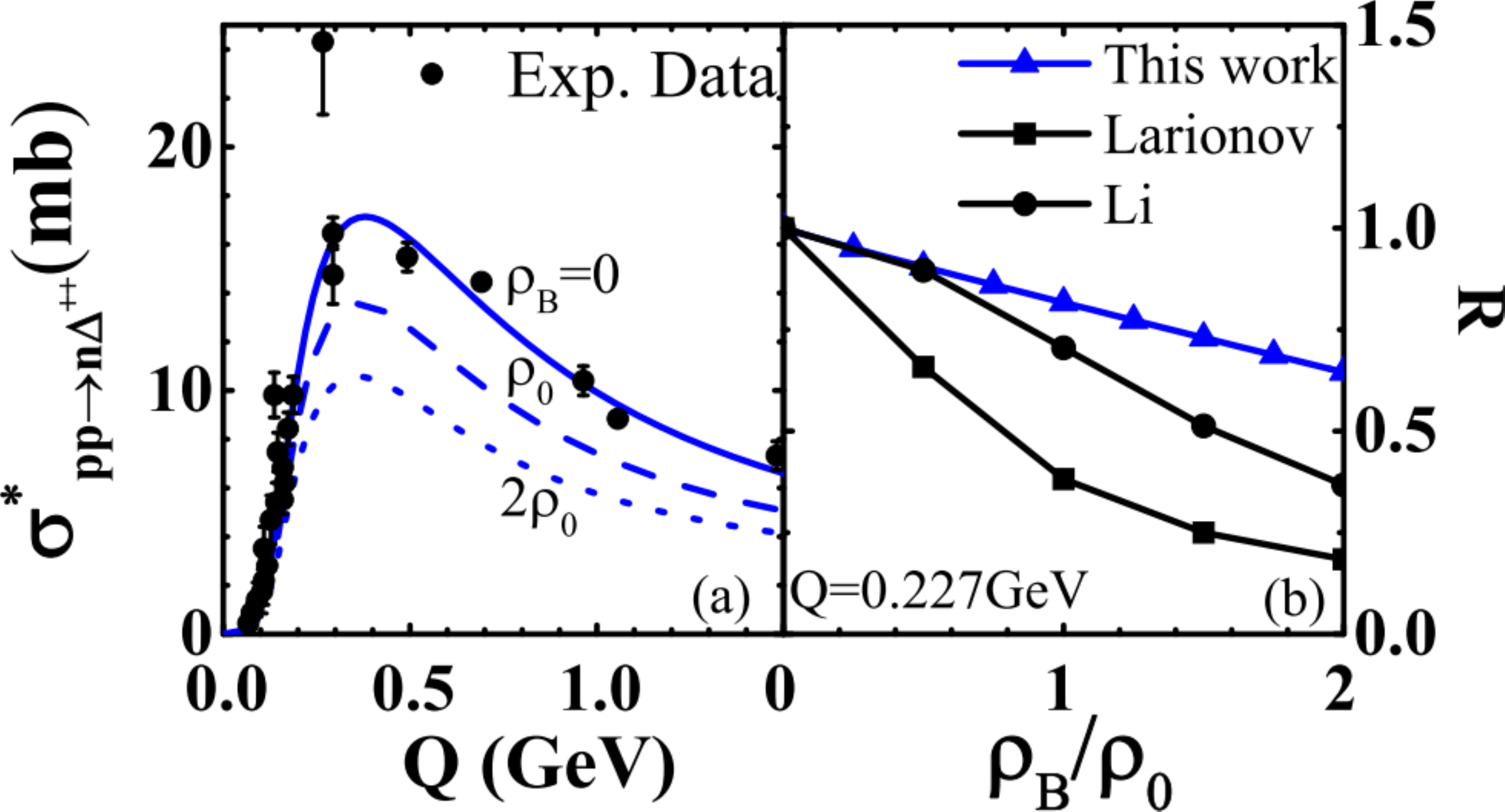}
    \caption{(Color online) (a)$\sigma^*_{pp\rightarrow n\Delta^{++}}$ as a function of $Q$ for $\rho_B$=0, $\rho_0$ and 2$\rho_0$, the experimental data are taken from Ref. \cite{Baldini1987};(b) $R=\sigma^*_{pp\rightarrow n\Delta^{++}}/\sigma^{\text{free}}_{pp\rightarrow n\Delta^{++}}$ as a function of density at $Q$=0.227 GeV ($E_{b}$=0.8 GeV). All of these results are for symmetric nuclear matter. The blue line with the triangles is the calculated results in this paper, and the black lines with the squares and circles correspond to the results from Refs. \cite{Larionov2003, QingfengLi2017}, respectively.}\label{fig3}
\end{center}
\end{figure}

Figure~\ref{fig3} (a) presents the $pp\rightarrow n\Delta^{++}$ cross section as a function of $Q$ for symmetric nuclear matter at $\rho_B=0$, $\rho_0$, and $2\rho_0$.  $Q$ is defined as
\begin{eqnarray}
\label{eq:qeff}
Q&=&\sqrt{s_{in}}-\sqrt{s_{\text{th}}}\\\nonumber
&=&E^*_{N_1}+E^*_{N_2}+\Sigma^0_{N_1}+\Sigma^0_{N_2}\\\nonumber
&&-m^*_{N_3}-m^*_{\Delta,\text{min}}-\Sigma^0_{N_3}-\Sigma^0_{\Delta}\\\nonumber
&\simeq &(E^*_{N_1}-m^*_{N_1})+(E^*_{N_2}-m^*_{N_2})\\\nonumber
&&+m_{N_1}+m_{N_2}-m_{N_3}-m_{\Delta,\text{min}}\\\nonumber
&&+\Delta\Sigma^S+\Delta\Sigma^0,
\end{eqnarray}
which represents the kinetic energy above the pion production threshold energy $\sqrt{s_{\text{th}}}=m^*_{N_3}+m^*_{\Delta,\text{min}}+\Sigma^0_{N_3}+\Sigma^0_\Delta$. The blue solid line represents the calculated $pp\rightarrow n\Delta^{++}$ cross section in free space. The black circles are the experimental data \cite{Baldini1987}. The calculation can well reproduce the data of $\sigma^{\text{exp}}_{NN\rightarrow N\Delta}$ except for the highest data point around $E_b$=1 GeV.
The dashed and dotted lines are the results for $\rho_B=\rho_0$ and $\rho_B=2\rho_0$, respectively. Comparing with the free space $pp\rightarrow n\Delta^{++}$ cross section, the in-medium $pp\rightarrow n\Delta^{++}$ cross sections in symmetric nuclear matter decrease with the density increasing for all channels. This is because the elementary two-body cross section $\tilde{\sigma}^*(m_\Delta^*)$ decreases with the reduction of $m_N^*$ and $m_{0,\Delta}^*$, which is consistent with the results in Ref. \cite{QingfengLi2017}.
The in-medium cross section for $nn\rightarrow p\Delta^{-}$ is equal to $pp\rightarrow n\Delta^{++}$, and other channels can be obtained based on the product of the isospin Clebsch-Gordan coefficients, which is 1/3 of $\sigma^*_{pp\rightarrow n\Delta^{++}}$.

In Fig.~\ref{fig3} (b), we show the medium correction factor $R=\sigma^*_{NN\rightarrow N\Delta}/\sigma^{\text{free}}_{NN\rightarrow N\Delta}$ as a function of density for Q=0.227 GeV($E_{b}$=0.8 GeV). The blue line is the result obtained in this paper. For the symmetric nuclear matter, all the channels have the same reduction factor. The $R$ values decrease with the density increasing, this behavior is consistent with the theoretical results from the  RBUU (line with the circles) \cite{QingfengLi2017} and the one-pion exchange model \cite{Larionov2003} (line with the squares), and this behavior has also been verified in the calculations of  transport models \cite{Larionov2003,Song2015} for reproducing the pion yield data \cite{Pelte97,Resid07}, whereas the reduction in our case is smaller than that obtained in the transport model calculations\cite{Song2015}. It may hint that the form factor in the Lagrangian needs to consider the in-medium correction.

\begin{figure}[htbp]
\begin{center}
    \includegraphics[scale=0.4]{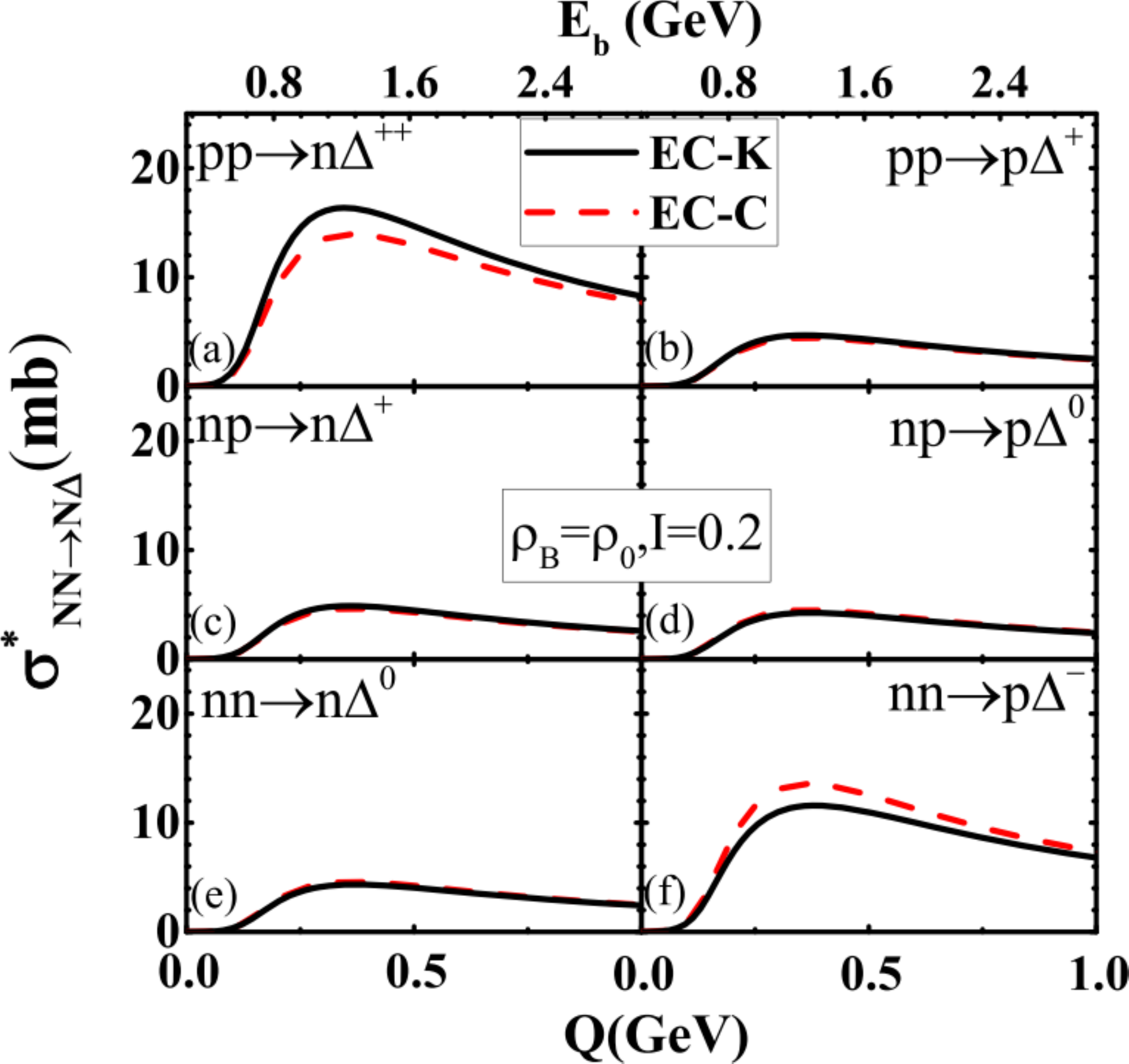}
    \caption{(Color online)  $\sigma^*_{NN\rightarrow N\Delta}$ for different channels at $I=0.2$, obtained with the EC-K (black solid lines)  and the EC-C  at (red dashed lines) $\rho_B=\rho_0$.}\label{fig4}
\end{center}
\end{figure}

Figures~\ref{fig4} (a)-~\ref{fig4}(f) present the results of $\sigma^*_{NN\rightarrow N\Delta}$ at $\rho_0$ with the isospin asymmetry $I$=0.2 for six channels, i.e. $pp\rightarrow n\Delta^{++}$, $pp\rightarrow p\Delta^{+}$, $np\rightarrow n\Delta^{+}$, $np\rightarrow p\Delta^{0}$, $nn\rightarrow n\Delta^{0}$ and $nn\rightarrow p\Delta^{-}$, respectively.
The red dashed lines are the results obtained with the EC-C, and the black solid lines are for the EC-K. For the $pp\rightarrow n\Delta^{++}$ channel, the cross section obtained with the EC-C is reduced relative to that with the EC-K; but for  $nn\rightarrow p\Delta^-$, the cross section obtained with the EC-C is enhanced relative to that with the EC-K. This can be simply understood from the isospin effects on  $Q$ which is the input of the formula of the  in-medium cross section. In the case of the EC-C,  $Q$ has an additional term $\Delta\Sigma^0$ compared to the EC-K where the isospin effects mainly come from the difference in $\Delta\Sigma^S$. For example, for the $pp\rightarrow n\Delta^{++}$ channel, the $Q$ values are reduced for the EC-C relative to that in the EC-K due to $\Delta\Sigma^0<0$. Thus, a larger $E^*-m^*(\approx p^2/2m^*)$ is needed for the  EC-C than that for the EC-K at a given $Q$ value by using Eq. (~\ref{eq:qeff}). This effect is similar to the decrease in the effective mass which will result in the reduction of the in-medium $pp\rightarrow n\Delta^{++}$ cross section relative to that for the EC-K. For  $nn\rightarrow p\Delta^-$, $\Delta\Sigma^0>0$ and the opposite behavior can be observed.
Consequently, the medium correction factor $R$ will be influenced as well.

\begin{figure}[htbp]
\begin{center}
\includegraphics[scale=0.41]{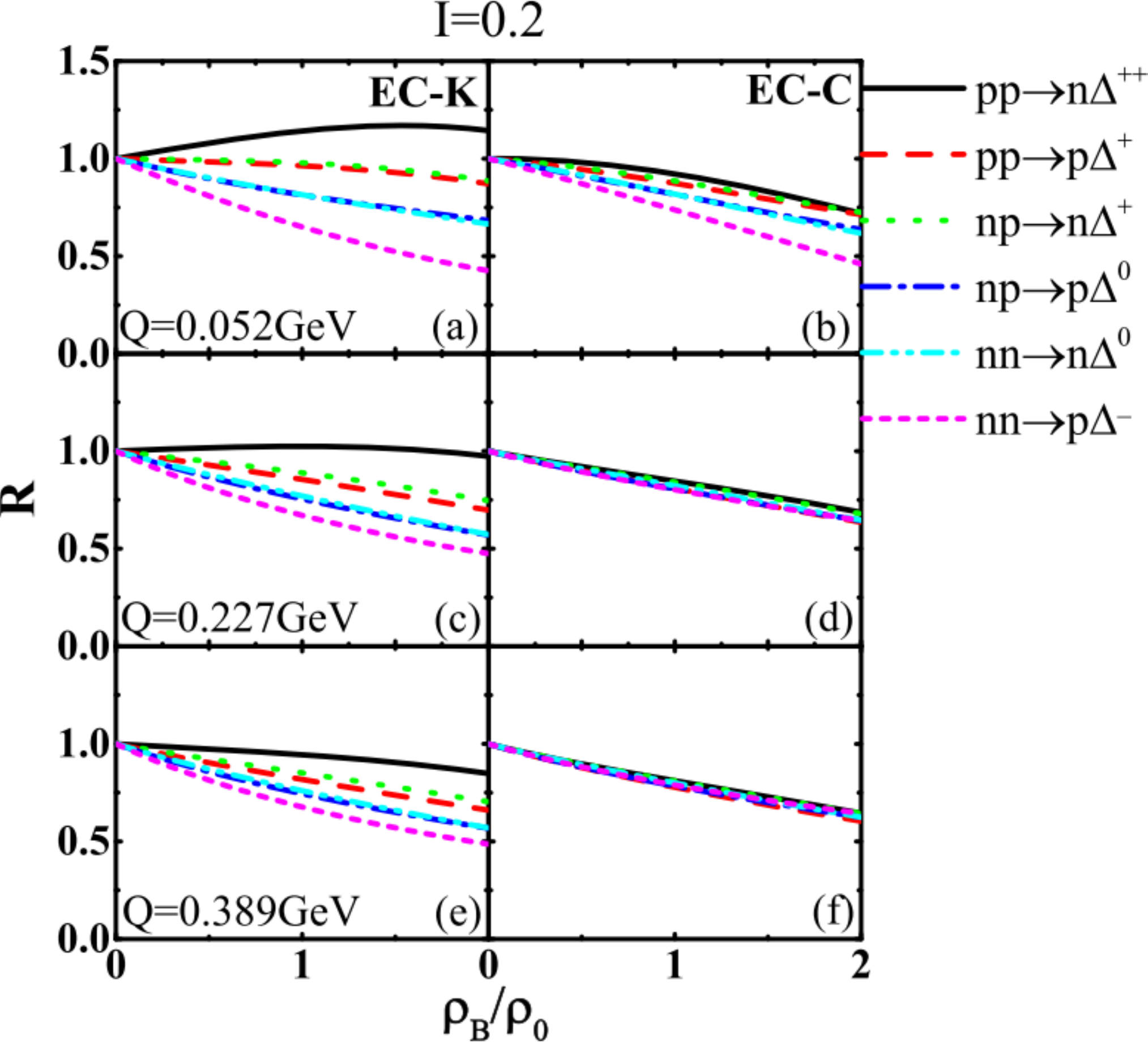}
\caption{(Color online) The medium correction factor $R$ as a function of density for different channels (with different colors),  for the beam energy at $Q$= 0.052, 0.227 and 0.389 GeV ($E_b=0.4$, 0.8 and 1.2 GeV) in isospin asymmetric matter at $I=0.2$. The left panels are for the EC-K, and the right panels are for the EC-C.}\label{fig5}
\end{center}
\end{figure}

In Fig. ~\ref{fig5}, we present $R$ as a function of density at $I=0.2$ for different channels.
The upper, middle, and bottom panels correspond to the results for $Q$=0.052 ($E_b$=0.4 GeV), 0.227 ($E_b$=0.8 GeV), and 0.389 GeV($E_b$=1.2 GeV), respectively. The
left panels are the results for the EC-K, and the right panels are the results for the EC-C.
As shown in the left panels of Fig.~\ref{fig5}, the $R$ ratios for different channels are split clearly at all the energies we analyzed, i.e., the $R$ values strongly depend on the charge state of $NN\rightarrow N\Delta$. The order of $R$ for different channels is $R(pp\rightarrow n\Delta^{++})>R(Np\rightarrow N\Delta^{+})>R(Nn\rightarrow N\Delta^{0})>R(nn\rightarrow p\Delta^{-})$ which is  the same as the order of the $\Delta$ effective mass and is consistent with the result in Ref. \cite{QingfengLi2017}.

In the case of the EC-C, the $R$ ratios for all channels decrease as a function of density. Near the threshold energy, the $R$ values clearly depend on the channels of $NN\rightarrow N\Delta$ and $R(pp\rightarrow n\Delta^{++})>R(Np\rightarrow N\Delta^{+})>R(Nn\rightarrow N\Delta^{0})>R(nn\rightarrow p\Delta^{-})$, but the magnitude of $R$ splitting between different channels becomes weaker than that for the EC-K because $\Delta\Sigma^0$ can wash out the isospin effects from $\Delta\Sigma^S$. With the beam energy increasing up to 0.8 GeV, the splitting of $R$ between the different channels of $NN\rightarrow N\Delta$ become smaller. Especially, the difference in $R$ between different channels tends to vanish above 1.2 GeV. This conclusion is different from the prediction in Ref. \cite{QingfengLi2017} where they still found the obvious difference in $R$ on different channels around 1 GeV. The reason is that the isospin splitting effects on $R$ mainly come from the isospin splitting of $\Delta$ effective masses in the case of the EC-K, which has been verified in the left panels of Fig.~\ref{fig5} and Ref. \cite{QingfengLi2017}. However,  $\Delta\Sigma^0$ can give the opposite contributions to $Q$ through the vector self-energy when one properly considers the energy conservation for incoming and outgoing particles. It leads to the reduction of isospin effects caused by the isospin splitting of the effective mass. When the beam energy is high enough, the contributions from the potential energy become smaller relative to the kinetic-energy part, and the isospin splitting of $R$ tends to vanish.
It implies that adopting the isospin channel-independent $R$ in the transport models is reasonable at the energy above 1 GeV, but our results further confirm that the channel dependence of $R$ should be taken into account near the threshold energy.

\section{Summary}
\label{summary}
To summarize, we have studied the isospin-dependent in-medium $NN\rightarrow N\Delta$ cross section in isospin asymmetric nuclear matter within the one-boson exchange model by including the $\delta$ and $\rho$ mesons. As a short living resonance, a parameterization formula of $\Delta$ mass distribution is involved in the calculation of $\sigma^*_{NN\rightarrow N\Delta}$.
With the proper energy conservation in asymmetric nuclear matter, our results confirm that the $\sigma^*_{NN\rightarrow N\Delta}$ are suppressed relative to the cross section in free space, and the medium correction factor $R$ also depends on the channels of the $NN\rightarrow N\Delta$ process near the threshold energy. However, the isospin splitting of $R$ becomes weaker at the beam energy above 0.8 GeV because the changes in scalar and vector self-energies become smaller relative to the kinetic-energy part. Our paper provides a theoretical information of the isospin-dependent medium correction factor $R$, which will be very useful for the further developing  isospin-dependent transport models.

Furthermore, the medium effect on pion could modify the $\Delta$ width and its mass distribution near the threshold \cite{xujun10}, and thus the theoretical work in this aspect on the in-medium $NN\rightarrow N\Delta$ cross section should also be investigated in the future. It will extend our understanding of the isospin dynamics in heavy-ion collisions.

\acknowledgments
The authors acknowledge helpful discussions with Professor J. Xu and Professor Q. Li. This work has been supported by the National Natural Science Foundation of China under Grants No.11875323, 11875125, 11475262, 11365004, 11375062, 11790323, 11790324 and 11790325 and the National Key R\&D Program of China under Grant No.\ 2018 YFA0404404.







\end{document}